\documentclass[usenatbib]{mn2e}
\usepackage[dvipdfm]{graphicx,color}
\usepackage{amssymb}
\usepackage{txfonts}
\usepackage{url}

\newcommand{\wco}{{W_\mathrm{CO}}}
\newcommand{\msun}{{\rm M}_{\sun}}

\newcommand\src{G73.9+0.9}
\newcommand{\g}{$\gamma$}
\newcommand{\fermi}{{\textit{Fermi}}}
\newcommand{\planck}{{\textit{Planck}}}
\newcommand{\xmm}{{\textit{XMM-Newton}}}

\newbox\grsign \setbox\grsign=\hbox{$>$} \newdimen\grdimen \grdimen=\ht\grsign
\newbox\simlessbox \newbox\simgreatbox \newbox\simpropbox
\setbox\simgreatbox=\hbox{\raise.5ex\hbox{$>$}\llap
     {\lower.5ex\hbox{$\sim$}}}\ht1=\grdimen\dp1=0pt
\setbox\simlessbox=\hbox{\raise.5ex\hbox{$<$}\llap
     {\lower.5ex\hbox{$\sim$}}}\ht2=\grdimen\dp2=0pt
\setbox\simpropbox=\hbox{\raise.5ex\hbox{$\propto$}\llap
     {\lower.5ex\hbox{$\sim$}}}\ht2=\grdimen\dp2=0pt
\def\ga{\mathrel{\copy\simgreatbox}}
\def\la{\mathrel{\copy\simlessbox}}
\def\simprop{\mathrel{\copy\simpropbox}}

\title[High-energy gamma-ray detection of G73.9+0.9]{The high-energy gamma-ray detection of G73.9+0.9,\\ a supernova remnant interacting with a molecular cloud} 
\author[A. Zdziarski et al.]{Andrzej A. Zdziarski,$^1$\thanks{E-mail: aaz@camk.edu.pl (AAZ); Denys.Malyshev@unige.ch (DM);
wilhelmi@ieec.uab.es (EOW); giovanna.pedaletti@desy.de (GP); Ruizhi.Yang@mpi-hd.mpg.de (RY)} Denys Malyshev,$^{2\star}$ Emma de O\~na Wilhelmi,$^{3\star}$\newauthor  Giovanna Pedaletti,$^{4,3\star}$ Ruizhi Yang,$^{5\star}$ Maria Chernyakova,$^{6,7}$ Marcos L{\'o}pez-Caniego,$^8$\newauthor Joanna Miko{\l}ajewska$^1$  and Rupal Basak$^1$\\
$^{1}$Centrum Astronomiczne im.\ M. Kopernika, Bartycka 18, PL-00-716 Warszawa, Poland\\
$^2$ISDC Data Centre for Astrophysics, Astronomy Department, Geneva University, Ch. d'Ecogia 16, Versoix 1290, Switzerland\\
$^{3}$Institut de Ci\`encies de l'Espai (IEEC-CSIC), Campus UAB, Torre C5, 2a planta, 08193 Bellaterra, Spain\\
$^4$Deutsches Elektronen-Synchrotron (DESY), D-15738 Zeuthen, Germany\\
$^{5}$Max-Planck-Institut f\"ur Kernphysik, P.O. Box 103980, 69029
Heidelberg, Germany\\
$^6$School of Physical Sciences, Dublin City University, Glasnevin, Dublin 9, Ireland\\ 
$^7$DIAS, Fitzwiliam Place 31, Dublin 2, Ireland\\
$^8$European Space Agency, ESAC, Planck Science Office, Camino bajo del Castillo, s/n, Urbanizaci\'{o}n Villafranca del Castillo, Villanueva de la Ca\~{n}ada,\\ Madrid, Spain
}

\begin{document}

\date{Accepted 2015 September 16.  Received 2015 September 14; in original form 2015 July 16}

\pagerange{\pageref{firstpage}--\pageref{lastpage}} \pubyear{2015}

\maketitle

\label{firstpage}

\begin{abstract}
We have analysed the \fermi\/ LAT data on the SNR \src. We have confirmed a previous detection of high-energy \g-rays from this source at a high significance of $\simeq 12\sigma$. The observed spectrum shows a significant curvature, peaking in $E F_E$ at $\sim$1 GeV. We have also calculated the flux upper limits in the mm-wavelength and X-ray ranges from \planck\/ and \xmm, respectively. We have inspected the intensity of the CO (1$\rightarrow $0) emission line and found a large peak at a velocity range corresponding to the previously estimated source distance of $\sim$4 kpc, which may indicate an association between a molecular cloud and the SNR. The \g-ray emission appears due to interaction of accelerated particles within the SNR with the matter of the cloud. The most likely radiative process responsible for the \g-ray emission is decay of neutral pions produced in ion-ion collisions. While a dominant leptonic origin of this emission can be ruled out, the relativistic electron population related to the observed radio flux will necessarily lead to a certain level of bremsstrahlung \g-ray emission. Based on this broad-band modelling, we have developed a method to estimate the magnetic field, yielding $B\ga 80\,\mu$G at our best estimate of the molecular cloud density (or less at a lower density). \src\ appears similar, though somewhat weaker, to other SNRs interacting with a local dense medium detected by the LAT. 
\end{abstract}

\section{Introduction}
\label{intro}

G73.9+0.9 is a supernova remnant (SNR) located in the Cygnus arm of the Galaxy. It was first observed and identified as an SNR in radio in the Effelsberg Galactic Plane survey \citep{reich86}. High-resolution imaging at 1.42 GHz (see \citealt{kothes06} and references therein) shows a morphology consisting of a shell-like feature to the south with a sharp outer boundary and a centrally peaked diffuse emission to the east. 

\begin{figure*}
 \centering  
  \includegraphics[width=1.3\columnwidth]{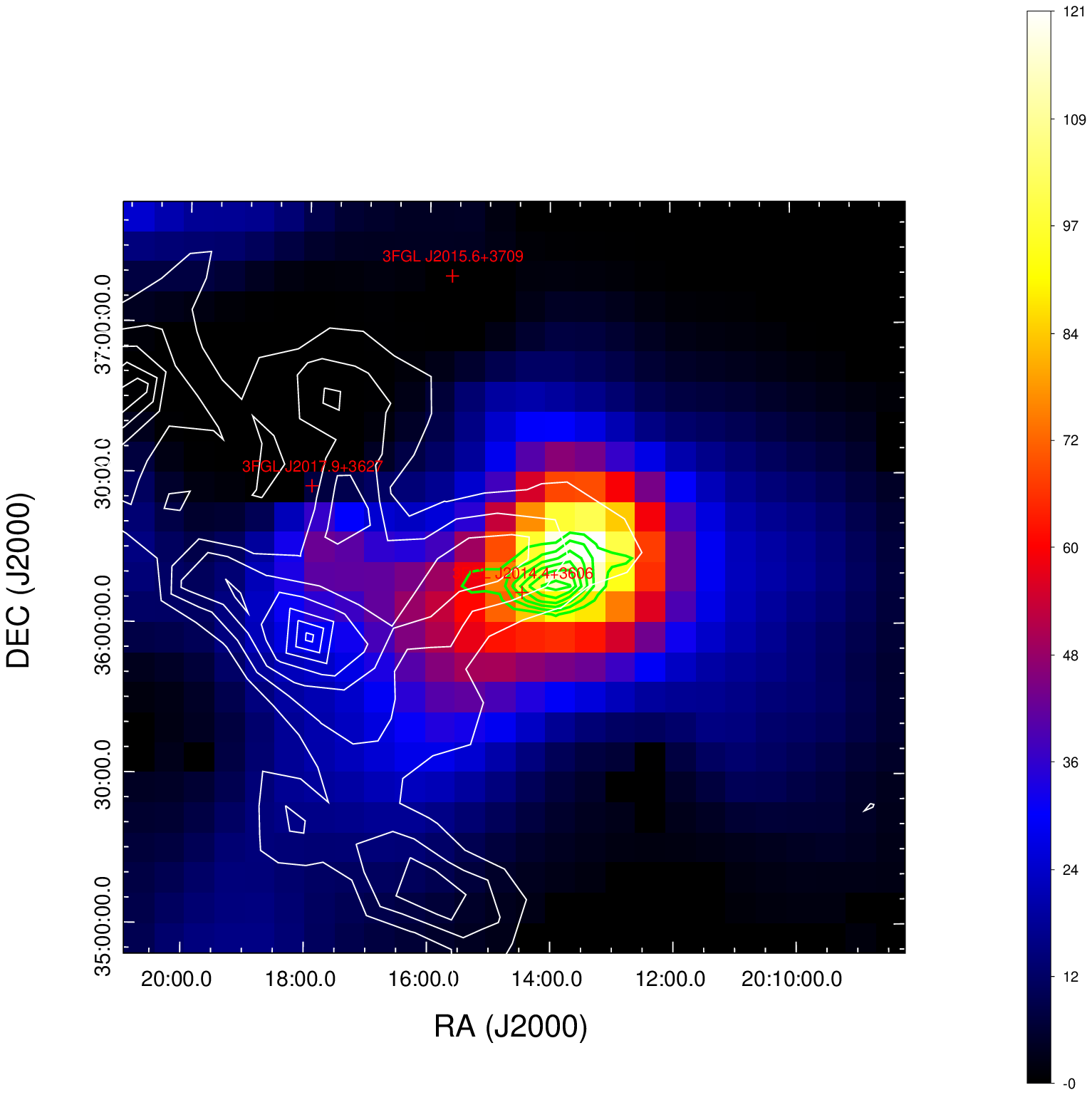}
  \caption{The colour image shows the $T_{\rm s}$ map of the region in 0.3--3\,GeV energy band with $0.1\degr \times 0.1\degr$ pixels.  All sources except 3FGL J2014.4+3606 (\src) were included into the region model and thus are not present on the map; $\sqrt{T_{\rm s}}$ gives the significance of a point-like source added to the model at a selected point. The green contours show the image of the SNR in 4.85 GHz continuum emission \citep{GB6survey}. The white contours are derived from the intensity data of the 115 GHz CO line (1$\rightarrow $0; \citealt{newdame_co}) in the 1--7 km s$^{-1}$ velocity range.}
 \label{image}
\end{figure*}

The radio emission has the total flux of $\sim$8 Jy at 1.4 GHz and the energy index of $\alpha\simeq -0.23$ ($F_\nu\propto \nu^\alpha$; measured between 0.4 and 10 GHz), corresponding to the photon index of $\Gamma=1.23$ ($F_\nu\propto \nu^{1-\Gamma}$). \citet*{lorimer98} searched for pulsed emission from the centre of the shell using the 76-m Lovell radio telescope, but without success. 

The region towards the source has been observed by the \planck\/ mission\footnote{\url{wiki.cosmos.esa.int/planckpla2015/index.php/Main_Page}} \citep{tauber10,plancka}, but since \src\ lies in a rather complex region of the Galactic plane, no results have been reported. Here, we obtain upper limits in the 30--143 GHz range, see Section \ref{fermi}.

Optical line diffuse emission has been observed towards the direction of this SNR \citep*{lozinskaya93,mavromatakis03}. The optical spectrum implies an electron density of $n_{\rm e}\la 50$ cm$^{-3}$ at some locations within the SNR, and a moderate shock velocity of $v_{\rm sh}\la 90$ km s$^{-1}$. These observations have allowed an estimate of the age of the SNR to be $T\sim 11$--12 kyr. Its distance has been variously estimated. The most recent determination of \citet{pavlovic13} puts it $D=4.0$ kpc, based on the radio surface-brightness to diameter relation. The radial velocity of \src\ has been measured by \citet{lozinskaya93} as $4\pm 3$ km s$^{-1}$. 

The SNR has not been detected in X-rays. We obtain an upper limit from \xmm, see Section \ref{fermi}.  The region has also been observed with Imaging Cherenkov Telescopes at very high energies, but no signal has been found. In particular, it was covered as part of the HEGRA Galactic Plane survey \citep{hegra_scan}. The 2-h exposure resulted in an upper limit on the number flux of $N(>\!\! E)<3.15\times 10^{-12} (E/\mathrm{1\, TeV})^{-1.59}\, \mathrm{cm}^{-2}\, \mathrm{s}^{-1}$ \citep{hegra_scan}. The VERITAS array also observed this region; the preliminary upper limit at the location of the SNR is $N(>\!300\,{\rm GeV})\la 2.3\times 10^{-12}\, \mathrm{cm}^{-2} \,\mathrm{s}^{-1}$ \citep{veritas_cygnus}.

At high-energy \g-rays (0.1 GeV$\la E\la $100 GeV), \citet*{Malyshev13} have reported a detection of \g-ray emission from the direction of G73.9+0.9 using the Large Area Telescope (LAT) on board of \fermi. The \g-ray emission was discovered when investigating the region around Cyg X-1, which lies $3.2\degr$ away of \src. The \g-ray source was detected with the test statistic ($T_{\rm s}$, \citealt{mattox96}) of $\simeq$50, at RA 303.42, Dec.\ 36.21, which is only $\sim\! 10'$ away from the catalogue position of the remnant, at RA 303.404 (20 13 37), Dec.\ 36.115 (36 06 54.0), corresponding to the Galactic coordinates of $l = 73.77$, $b = 0.94$. This detection has recently been confirmed in the \fermi\/ Large Area Telescope Third Source Catalogue (3FGL, \citealt{3rdcat}), as the source named J2014.4+3606 with the significance\footnote{\url{heasarc.gsfc.nasa.gov/W3Browse/fermi/fermilpsc.html}} of $4.5\sigma$. Here, we re-analyse a significantly larger data set with the present instrumental calibration, and obtain significantly better statistical results. 

We present here a dedicated analysis of the currently available data set obtained with the LAT on the SNR and an investigation of the molecular content around it, derived from an analysis of the CO line (1$\rightarrow $0) intensity at 115 GHz (2.6 mm) in the region, observed with the CfA telescope \citep{dame_co}. We also study models possibly explaining the multi-wavelength spectrum of \src.

\section{Multi-Wavelength data analysis and results}
\label{multi}

\subsection{The imaging and spectral analysis}
\label{fermi}

We have analysed the \fermi\/ LAT data derived from a $15\degr$ radius region centred on the position of \src. Almost seven years of data obtained during MJD\,54682--57209 (2008 Aug.\ 04--2015 Jul.\ 06) were processed using the LAT standard tools (software version \texttt{v10r0p5}), and analysed with the \texttt{P8R2} response functions (\texttt{SOURCE} class photons). We have applied the standard cleaning to suppress the effect of the Earth albedo background, excluding time intervals when the Earth was in the field of view (FoV; when its centre was $>52\degr$ from the zenith), and those in which a part of the FoV was observed with the zenith angle $>90\degr$. 

We have selected events with energies between 0.1 GeV to 300 GeV, performing binned likelihood analysis. It is based on the fitting of a model of diffuse and point source emission within selected region to the data. The spatial model includes diffuse Galactic and extragalactic backgrounds (\texttt{gll\_iem\_v06.fits} and \texttt{iso\_P8R2\_SOURCE\_V6\_v06.txt} templates) and the sources from the 3FGL catalogue. We split the whole energy range into narrow energy bins, performing the fitting procedure in each bin separately. In each bin, the spectral shapes of all sources are assumed to be power laws with $\Gamma=2$. The spectral shapes of diffuse Galactic and extragalactic backgrounds are given by the corresponding templates. The normalisations of the fluxes of all sources and the diffuse background are treated as free parameters during the fitting. The analysis is performed with the python tools\footnote{\url{fermi.gsfc.nasa.gov/ssc/data/analysis/scitools/python_tutorial.html}}. The upper limits are calculated with the \texttt{UpperLimits} module provided with the \fermi/LAT software and correspond to the 95 per cent ($\simeq 2.5\sigma$) false-chance probability.

We localise the position of \src\ by building the $T_{\rm s}$ map of the selected region. The colour image in Fig.~\ref{image} shows this map, which illustrates the significance of adding a point source to the model of the region, which follows the $\chi^2$ distribution and is approximately $\simeq\! \sqrt{T_{\rm s}}$ \citep{mattox96}. The model in this case includes all the sources described above, except 3FGL~J2014.4+3606, which corresponds to \src\ in the 3FGL catalogue. We see a clear \g-ray excess at the position of \src\ in the 1--3~GeV energy band. The centroid of the image is the same as that in the 3FGL catalogue. The source shows a point-like morphology, with the $3\sigma$ upper limit on its radius (corresponding to the decrease of $T_{\rm s}$ by 9 for the uniform-brightness disc model by 9 with respect to the point-source model) of $0.23\degr$ in the 0.3--3 GeV energy band. The maximum of the $T_{\rm s}$ for the \g-ray source lies $0.1\degr$ from the centre of the radio emission from \src, which offset is not statistically significant. This is also shown in Fig.\ \ref{image}, where the black contours show the 4.85 GHz continuum emission from the GB6/PMN survey \citep{GB6survey}. 

\begin{figure}
 \centering  
  \includegraphics[width=\columnwidth]{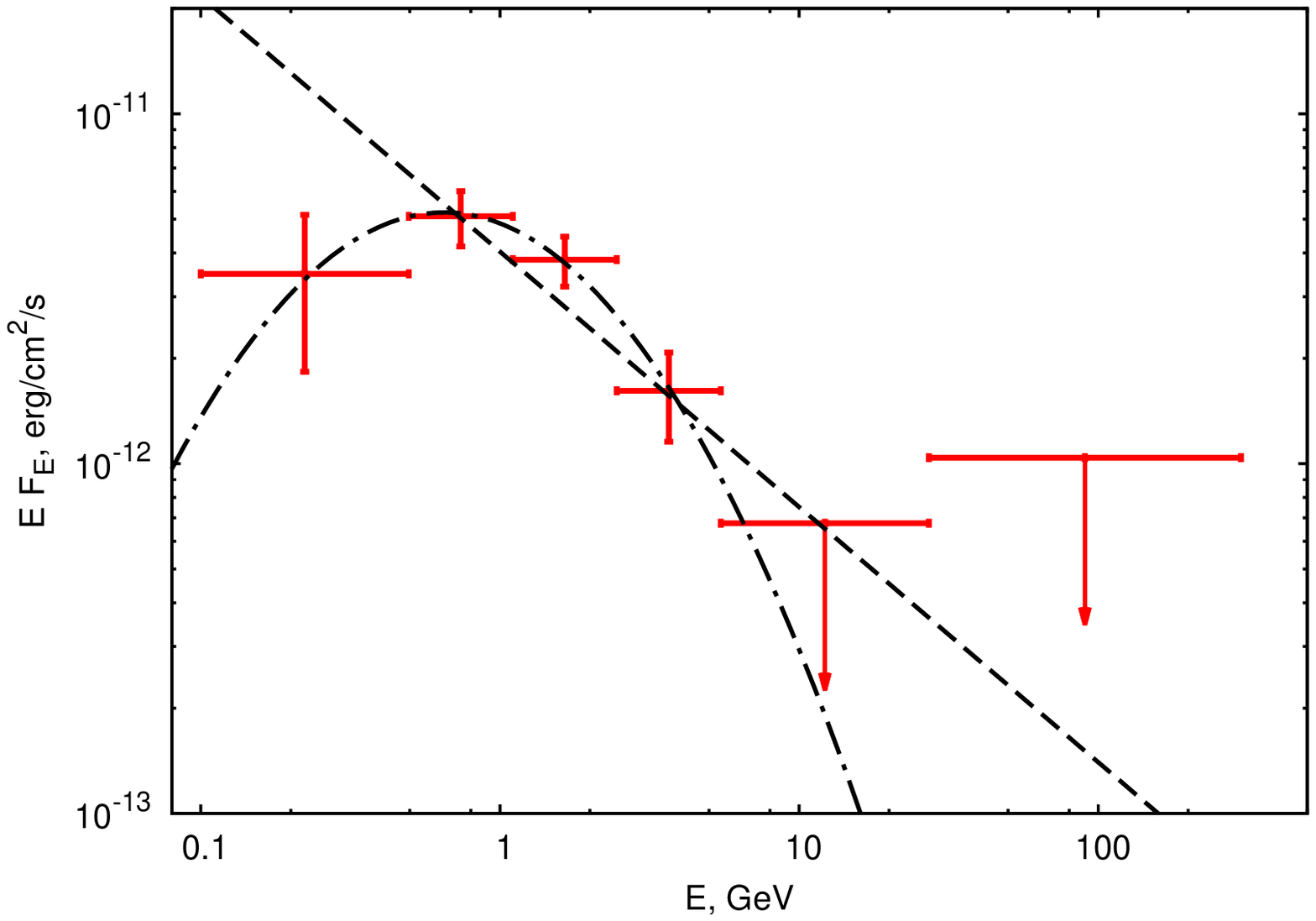}
  \caption{The $\gamma$-ray spectrum of a point-like source at the position of \src. The shown error bars present $1\sigma$ statistical uncertainties and the upper limits correspond to 95 per cent confidence (about $2.5\sigma$). The dashed and dot-dashed curves shows the best-fit power-law and log-normal spectrum, respectively. 
}
 \label{spectrum}
\end{figure}

The energy spectrum for a point-like source centred at the position of \src\ was derived by means of the binned likelihood fitting, as described above. We show it in Fig.\ \ref{spectrum}. The shown error bars correspond to $1\sigma$ statistical errors. In the energy range of 0.1--300~GeV the spectrum can be fitted with a power law, which we show in Fig.\ \ref{spectrum}, with the photon index of $\Gamma = 2.73 \pm 0.13$ and the normalization of $N_0 = (2.17 \pm 1.10)\times 10^{-11}$ erg cm$^{-2}$ s$^{-1}$, defined as the value of $E F_E$ at 0.1 GeV. This fit yields the $T_{\rm s}\simeq 64$, which corresponds to $\sigma\simeq 7.7$. This confirms, at a higher significance, the presence of this source in \g-rays discovered by \citet{Malyshev13}. 

However, the spectral points show a clear curvature in the spectrum, with the peak around 1 GeV. We confirm the high statistical significance of the presence of a curvature by fitting a log-normal model (see Appendix \ref{log}) to the 0.1--300 GeV photon-energy band, which yields a much higher $T_{\rm s}\simeq 150$, which corresponds to the significance of $\simeq 11.8\sigma$ (with an addition of one free parameter with respect to the power-law fit). The fit parameters are $N_0 = (7.7 \pm 1.0)\times 10^{-12}$ cm$^{-2}$ s$^{-1}$ MeV$^{-1}$ [defined as in equation (\ref{lognormal})], $E_{\rm b}=0.65\pm 0.04$ GeV, which is the peak of the log-normal distribution in $E F_E$, and $\beta=0.89\pm 0.07$. We also obtain a similar $T_{\rm s}$ for a broken power law fit, which has two more parameters than a single power law. We have also fitted the spectrum at $E\geq 1$ GeV by a power law, to facilitate comparison with \fermi\/ review papers, which give that index for other SNRs, see, e.g., \citet{brandt15}. We have obtained $\Gamma = 3.2 \pm 0.2$ and the normalization of $N_0 = (6.2 \pm 0.8)\times 10^{-12}$ erg cm$^{-2}$ s$^{-1}$, given as $E F_E$ at 1 GeV.

The integrated 1--100 GeV photon and energy fluxes derived directly from the obtained spectrum (i.e., treating the upper limits as zero points with the uncertainty treated as a systematic error) are $\simeq 2.0^{+1.9}_{-0.24} \times 10^{-9}$ cm$^{-2}$ s$^{-1}$ and $\simeq 5.2^{+5.0}_{-0.65} \times 10^{-12}$ erg cm$^{-2}$ s$^{-1}$, respectively. The estimated \g-ray luminosities for the 0.1--100, 0.5--100 and 1--100 GeV ranges are, respectively, $28.3^{+9.9}_{-5.9}$, $16.4^{+5.7}_{-1.9}$, and $9.9^{+6.7}_{-1.2} \times(D/4\,{\rm kpc})^2 \times 10^{33}$ erg s$^{-1}$. 

\begin{figure*}
 \centering  
  \includegraphics[width=13cm]{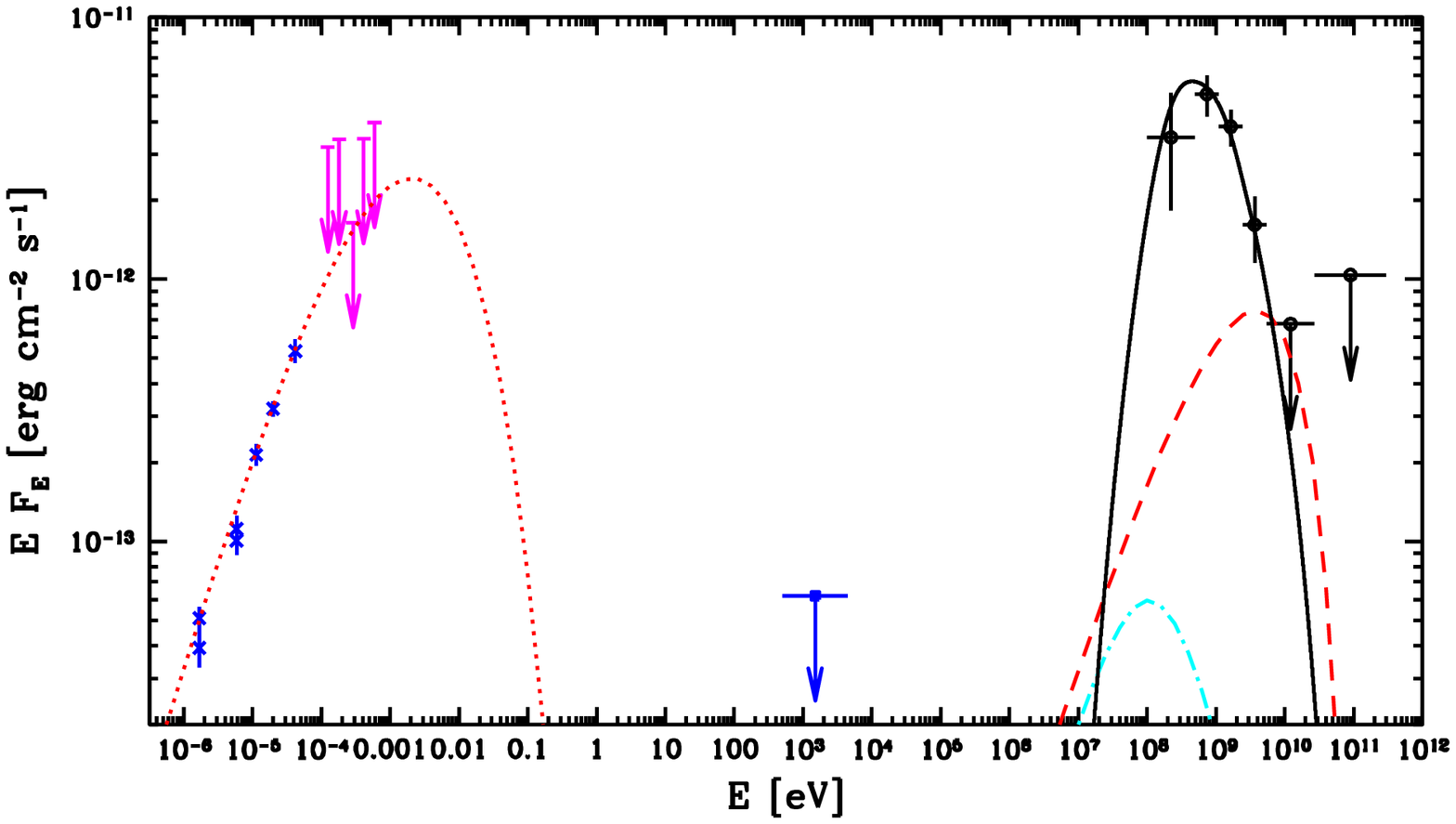}
  \caption{The observed broad-band spectrum of \src. The black symbols with open circles give our \g-ray measurements and upper limits, and the magenta symbols give our upper limits from \planck. The radio data from \citet{kothes06} and the upper limit from \xmm\ are shown by blue crosses and error bars, and by the blue upper limit with an open circle, respectively. The spectrum from decay of neutral pions in the hadronic model is shown by the black solid curve, see Section \ref{hadron}. The cyan dot-dashed curve shows the bremsstrahlung emission of e$^\pm$ from decay of charged pions in the same model assuming the SNR age of 12 kyr and at $n_{\rm e}=30$ cm$^{-3}$ and $B=80\,\mu$G. The corresponding synchrotron emission is below the shown range of fluxes. The radio spectrum is modelled by synchrotron emission (shown by the red dotted curve) from e-folded power law electrons with the minimum value of $B\gamma_0^2$ allowed by the radio data not showing any high-energy cutoff. The normalization of the spectrum yields $K_{\rm e} B^{(1+p)/2}/(4\upi D^2)$. The red dashed curve shows the maximum bremsstrahlung emission from the same electrons allowed by the data, which implies that $B\ga 80\,\mu$G. See Section \ref{electron} for details. 
}
 \label{SED}
\end{figure*}

We have also searched for X-ray emission based on the \xmm\/ pointed observation of the field of \src\ on 2006 Apr.\ 19 (ObsID 0301880601, full window mode, PI: O. Kargaltsev). The observation yields an upper limit of $\la 2.7\times 10^{-14}$ erg cm$^{-2}$ s$^{-1}$ in the energy range of 0.5--4.5 keV within a circle of $90\arcsec$ radius around the observation pointing of RA 303.48, Dec.\ 36.25. The analysis has been done using the \xmm\/ online upper-limit software\footnote{\url{www.ledas.ac.uk/flix/flix.html}}. The upper limit is from the MOS2 detector, for which the exposure time is 2550 s; the MOS1 and PN have much shorter exposures. From \xmm\/ and other X-ray observatories, there are no bright X-ray sources seen within the radius of $0.2\degr$ around the SNR centre. From the image of the source in radio at 1.42 GHz \citep{kothes06}, we have visually estimated the area of the peak of the SNR emission to be $\sim (0.1\degr)^2$. This yields an upper limit on the SNR emission within 0.5--4.5 keV of $<1.4\times 10^{-13}$ erg cm$^{-2}$ s$^{-1}$. This corresponds to $E F_E< 6.2\times 10^{-14}$ erg cm$^{-2}$ s$^{-1}$ (assuming $\Gamma=2$), shown in Fig.\ \ref{SED}. This upper limit has a systematic uncertainty related to the uncertain source area and the spectral slope, which we roughly estimate to be of a factor of three.

We have also searched for mm-wavelength emission using the \planck\/ 2015 data \citep{planckb,planckc}. The source does not appear in the Second \planck\/ Catalogue of Compact Sources \citep{planckd}. We have obtained $2\sigma$ upper limits in the 30--143 GHz frequency range using the 2015 full mission maps, and we show them in Fig.\ \ref{SED}.

The flux density of the source and the uncertainties used to calculate the upper limits have been measured as follows. For the flux density, aperture photometry centred at the position of the source has been used using an aperture with a radius equal to one FWHM. The flux densities have been corrected to take into account that a fraction of the beam solid angle falls outside the aperture. For the rms, we have used an annulus with the inner radius of 1 FWHM and the outer radius of 2 FWHM. For all the frequencies, the FWHMs are the effective beams as listed in column 3 of Table 2 of \citet{planckd}.

\subsection{A molecular cloud around \src}
\label{molecular}

Several GeV-bright SNRs have been shown to interact with co-located molecular clouds. The presence of such a cloud appears to be required for emission of \g-rays detectable by the LAT from old SNRs \citep{hewitt13}. A safe identification of interaction with such high density material would be a detection of masers (see, e.g., W51C, \citealt{w51c_maser}). A search in the direction of \src\ for both OH and H$_2$O masers, at 1.72 GHz and 22 GHz, respectively, gave no results \citep{ohmaser,hdueomaser}. Data from the I-GALFA H{\sc i} 21-cm line survey did not show any significant excess emission that can be associated with the remnant \citep{higalfa}. 

Therefore, in order to evaluate the molecular content at the location of the remnant, we use the intensity, $\wco$, of the CO line (1$\rightarrow $0) at 115 GHz (2.6 mm), given in units of K km s$^{-1}$. CO is the most commonly used tracer for molecular material, given the lack of a possibility to directly observe H$_2$. In order to obtain $N_\mathrm{H_2}$, the column density of
H$_{2}$, a conversion factor $X_\mathrm{CO}$ is then used
\citep*{xcoreview}. An extensive archive is provided by \citet{dame_co} and \citet*{newdame_co}. The data are released\footnote{\url{www.cfa.harvard.edu/rtdc/CO/}} in cubes of radial velocity (in the local standard of rest) and Galactic latitude and longitude. We can estimate the mass of the molecular material from $\wco$ as
\begin{equation}\label{eq:massco}
 M={m_\mathrm{H}\over X} D^2 \Delta\Omega_\mathrm{px} X_\mathrm{CO} {\sum_\mathrm{px}} \wco \propto N_\mathrm{H_2},
\end{equation}
where $X$ is the H mass fraction, $m_\mathrm{H}$ is the mass of the H nucleon, and $X_\mathrm{CO}$ is a conversion factor (see below), and $\Delta\Omega_\mathrm{px}$ corresponds to the solid angle subtended for each pixel in the map (square binning of $0.125\degr$ per side). The term $\sum_\mathrm{px} \wco$ takes into account the binning in velocity of the data cube and is obtained by summing the map content for the pixels in the desired sky region and desired velocity range and scaled by the bin size in velocity.

We have calculated the absolute value of the molecular mass using the conversion factor estimated by \citet{newdame_co}, $X_\mathrm{CO}= 1.8 \times 10^{20}\, \mathrm{cm}^{-2} (\mathrm{K\, km/s})^{-1}$. Studies of the estimation of the gas content using different tracers show compatible average results \citep{Pedaletti}. 

We have found that the velocity distribution of the CO content shows a clear excess in the range of velocity of $(-2.0,\,3.2)$ km s$^{-1}$. This overlaps with the velocity range measurement of $v\simeq 1$--7 km s$^{-1}$ of \citet{lozinskaya93}. Thus, we adopt the range of $v\simeq 1.0$--3.2 km s$^{-1}$ as the SNR systemic velocity. Based on it and using the Galacto-centric rotation-distance relation of \citet{clemens85}, we estimate the distance of \src\ of either 0.2--0.5 kpc or 4.3--4.5 kpc. These estimates bear some systematic uncertainties due to the uncertainty of the rotation curve of \citet{clemens85} and to the possible proper motion of the source. The lower range is highly unlikely, while the upper one approximately agrees with the estimate of 4.0 kpc of \citet{pavlovic13}. 

We have then estimated the mass content in a region of $0.15\degr$ sky integration radius around the position of the \g-ray source as $M \simeq 4300(D/4\,{\rm kpc})^{2}\msun$. The conversion of the mass to density is uncertain, given the lack of knowledge about the shape of the cloud. We provide a rough estimate of the electron density of the medium assuming a uniform spherical cloud with the radius equal to the sky integration radius, which is, obviously, a rather strong assumption. For $X=0.7$, this yields $n_{\rm e}\simeq 30(D/4\,{\rm kpc})^{-1}$ cm$^{-3}$. This agrees with the upper limit of 50\,cm$^{-3}$ of \citet{mavromatakis03}. However, the actual density can be lower if the radius of the cloud is higher than that corresponding to $0.15\degr$. Furthermore, we cannot exclude that the directional coincidence of the molecular cloud and the SNR is accidental. 

\subsection{The hadronic model for the $\mathbf{\gamma}$-ray spectrum}
\label{hadron}

Given the relatively high density of the medium where the SNR is located, two standard models can be considered to explain the observed \g-ray emission. In both, the emission is due to radiation of the particles accelerated in the supernova shock interacting with the dense medium around it. In the first model, the \g-rays are produced through the decay of neutral pions produced in the collisions between the accelerated ions at high energies and the ambient ions. In the second model, the \g-rays have a leptonic origin from non-thermal bremsstrahlung radiation of the accelerated non-thermal electrons on the ambient medium (see Section \ref{electron}). A possible contribution from inverse Compton radiation can be neglected at $\la 100$ GeV in comparison with the bremsstrahlung given the high ambient density we have measured (see, e.g., \citealt*{yang14}). 

We estimate here the hadronic \g-ray emission using the formalism in \citet{kafexhiu14}. The observed \g-ray spectrum is rather narrow, and we have found that a correspondingly narrow proton distribution is required to reproduce it. Given that we have only four data points, we have not attempted a formal fit, but only tested a range of distribution. The observed spectrum can be reproduced by a proton distribution with a power-law index of $p=2.0$ and an exponential cutoff at 20 GeV, i.e., $N(E)\propto E^{-2}\exp(-E/20\,{\rm GeV})$. Decay of $\pi^0$ produced in proton-proton collisions yields the spectrum shown by the solid curve in Fig.\ \ref{SED}. The energy content in the proton distribution above the threshold for pion production, i.e., $E> 0.3$ TeV, is $\simeq 4.7(D/4\,{\rm kpc})^3(n_{\rm e}/30\,{\rm cm})^{-1}\times 10^{48}$ erg. We note this is a modest energy content when comparing with the total kinetic energy of an SNR, even one with the age $\ga 10$ kyr. 

The steep high-energy part of the \g-ray spectrum is similar to those found in a number of GeV-bright SNRs and can be explained naturally by either a deficit of high-energy protons due to the cooling or escape of particles during inefficient cosmic ray acceleration in dense surrounding environments (see, e.g., \citealt*{aa96,malkov11,malkov12,gabici09}). Alternatively, it can be a result of the spectral shape emerging from re-acceleration of pre-existing cosmic rays in compressed clouds \citep{uchiyama10,bc82}.

We then consider the possible role of e$^\pm$ produced from decay of charged pions. We have calculated the rate of their production, $Q_{{\rm e}\pm}(\gamma)$, which is relatively similar both in shape and normalization to the photon production rate from $\pi^0$ decay. The rate $Q_{{\rm e}\pm}(\gamma)$ contains most of the energy around the Lorentz factors of $\gamma\sim (100$--2000). The e$^\pm$ lose then their energy in interactions with the background gas via Coulomb and bremsstrahlung, and with the magnetic field via the synchrotron process. We have calculated the energy loss rates, $\dot \gamma$, for those processes assuming $n_{\rm e}=30$ cm$^{-3}$ and $B=80\,\mu$G (see Section \ref{electron}). The time scale for the total energy loss, $\gamma/\dot \gamma$, is found to be $\ga 10^5$ yr for $\gamma\simeq (10^2$--$5)\times 10^4$, which is the range relevant for the e$\pm$ emission. This is much longer than the age of the SNR of $T\simeq 11$--12 kyr. Thus, the energy losses can be neglected and an upper limit on the current electron distribution within the source can be estimated as $N(\gamma)\simeq T Q_{{\rm e}\pm}(\gamma)$ at $T=12$ kyr. This is an upper limit since we neglected possible leakage of e$^\pm$ out of the source. We have then calculated the synchrotron and bremsstrahlung spectra for this $N(\gamma)$ (see Section \ref{electron} for the method). Both processes are found negligible. The latter component is shown by the dot-dashed curve in Fig.\ \ref{SED}, whereas the former is below the scale of the plot. 

Our present modelling predicts rather few photons at $E>50$ GeV. This can be tested by the CTA \citep{acero13}, the next generation of Cherenkov telescopes. If there is still emission at this photon energy range, the high angular resolution of the CTA would permit a detailed investigation of the acceleration region. 

\subsection{Constraints on the electron distribution and magnetic field from the broad-band spectrum}
\label{electron}

As we have found out above, the proton distribution required to explain the \g-ray spectrum is quite narrow. On the other hand, the observed lack of any visible curvature in the radio spectrum, see the black data points in Fig.\ \ref{SED}, requires the existence of an electron distribution with a power law form for at least two decades, though the 70-GHz \planck\/ upper limit requires some steepening by this frequency. This implies the protons and electrons are accelerated differently.

The local electron power-law index of $p\simeq 2\alpha+1\simeq 1.46$ is required. Then, the observational data impose constraints on the overall electron distribution and the magnetic field via the normalization of the radio spectrum (emitted via synchrotron), the lack of its curvature up to 10 GHz, and the \g-ray detection and upper limits (considering the contribution from bremsstrahlung by the same electrons). 

We first assume the distribution of relativistic electrons in the source to be a generalized e-folded power law, 
\begin{equation}
N(\gamma)=K_{\rm e}\gamma^{-p} \exp\left[- (\gamma/\gamma_0)^q\right].
\label{exp}
\end{equation}
We have found that either $q=1$ (exponential cutoff) or $q=2$ (super-exponential cutoff) lead to almost the same constraints on the magnetic field. Thus, we present the results only for $q=1$. We also consider electrons with a broken power-law shape, having the index of $p+\Delta p$ above $\gamma_0$. We assume a smooth transition between the two power-law parts,
\begin{equation}
N(\gamma)={K_{\rm e}\over \gamma^p +\gamma_0^{-\Delta p}\gamma^{p+\Delta p}}.
\label{bpl}
\end{equation}

Our procedure is as follows. The synchrotron power-law spectrum constrains its index and a combination of the electron normalization and the magnetic field strength, $B$. The presence or absence of a curvature constrains a combination of $B$ and $\gamma_0$. The resulting three parameters are
\begin{equation}
p,\quad {K_{\rm e} B^{(1+p)/2}\over 4\upi D^2}\equiv A_1\quad B\gamma_0^2\equiv A_2.
\label{parameters}
\end{equation}
The overall synchrotron spectrum depends only on these three parameters, with its shape given by $p$ and $A_2$, and the normalization, by $A_1$. The lack of any visible spectral curvature up to 10 GHz yields a lower limit of $A_2$. A further constraint can be provided by the bremsstrahlung spectrum emitted by the same electrons, given the density of the background matter and the observational constraints in \g-rays. This allows us to constrain all the four individual parameters, $p$, $B$, $K_{\rm e}$ and $\gamma_0$. 

Although the local power-law index of the electrons responsible for the observed radio emission has $p\simeq 1.46$ and the normalization of $A_1\simeq 1.3$ (cgs), a harder distribution is required in the presence of an exponential cutoff, equation (\ref{exp}) with $q=1$, which cutoff softens the local spectrum. We constrain the considered range of the index to $p\geq 1.2$, given the complete lack of harder distributions in radio SNRs \citep{kothes06}. At $p=1.2$ and $A_1\simeq 0.7$ (cgs), $A_2\simeq 5\times 10^4$ G corresponds to the (approximate) maximum spectral curvature required by the data at $\leq 10$ GHz and the minimum one required by the 70-GHz $2\sigma$ upper limit.

We show the synchrotron spectrum corresponding to those constraints (with the minimum of $B\gamma_0^2$) by the red dotted curve Fig.\ \ref{SED}. Since we calculate the emission from the entire SNR, we use the formula for the synchrotron emissivity averaged over the pitch angle \citep*{cs86,ggs88}, which we then integrate over the electron distribution. For a value of $A_2$ higher than the above lower limit, the synchrotron spectrum in the observed radio range will be flatter, and the peak in $\nu F_\nu$ will be higher and at a higher energy than that plotted. 

We then calculate the bremsstrahlung spectrum from the same electrons, using our estimated density of the molecular cloud, $n_{\rm e}\simeq 30$\,cm$^{-3}$. We use the bremsstrahlung formulae for $E\ga 0.1$ GeV, at which energies the contributions from ions and electrons have the same form \citep*{strong00}. The bremsstrahlung spectrum is $\propto K_{\rm e}n_{\rm e}(3+X)(1+X)^{-1}(4\upi D^2)^{-1}$. Its photon index at $E\ll \gamma_0 m_{\rm e}c^2$ is $\simeq p$, and the peak of its $E F_E$ spectrum is around $E\sim m_{\rm e}c^2 \gamma_0$ at roughly $E F_E \simprop n_{\rm e} K_{\rm e}\gamma^{2-p}$. We then compare the calculated bremsstrahlung spectrum to the constraints from the \fermi\/ observations. We find the \g-ray observations constrain $B\ga B_{\rm min}\simeq 80(D/4\,{\rm kpc})^{-2/(p+1)}\,\mu$G. Then $K_{\rm e}/(4\upi D^2)\simeq A_1/B^{(1+p)/2}$ and $\gamma_0\ga \gamma_{0,{\rm min}}\simeq (A_2/B)^{1/2}\simeq 2.5\times 10^4$. The model with $B_{\rm min}$ and $\gamma_{0,{\rm min}}$ is plotted in Fig.\ \ref{SED} as the red dashed curve. We see the predicted bremsstrahlung component is at the observational upper limit. We have also found that if we decrease (increase) $\gamma_0$ via a decrease (increase) of $A_2$, we need to harden (soften) the electron distribution in order to have the synchrotron spectrum passing through the data points. While the decrease of $A_2$ decreases the resulting bremsstrahlung spectrum, the decrease of $p$ increases it, and the value of $B_{\rm min}$ remains almost unchanged. Thus, our constraint is robust, independent of details of the estimate of the possible cutoff in the synchrotron spectrum.

We then repeated the calculations for electrons with a broken power-law distribution, equation (\ref{bpl}). However, we have found that unless $\Delta p\ga 1.5$, the resulting constraints impose an even higher $B_{\rm min}$. At $\Delta p=1.5$ (and $n_{\rm e}\simeq 30$\,cm$^{-3}$), $B_{\rm min}\simeq 80\,\mu$G, i.e., the same as for the exponential cutoffs. Thus, our constraint on $B$ is again robust, approximately independent of the assumed break or cutoff in the electron distribution.

However, the actual matter density within the SNR can be lower than 30\,cm$^{-3}$, e.g., due to the cloud length along the line of sight being more than we assumed. Then, the bremsstrahlung emission will be weaker, and the minimum allowed magnetic field strength will be lower. A relatively strict lower limit on the density is provided by that of the ISM, which we take as $n_{\rm e}= 1$\,cm$^{-3}$. In this case and for equation (\ref{exp}) with $q=1$, we find $B_{\rm min}\simeq 9\,\mu$G and $\gamma_{0,{\rm min}}\simeq 7.5\times 10^4$.

We point out that it would be of significant interest to detect \src\ at a frequency higher than the current detection at 10 GHz, to improve the constraint on the electron distribution obtained in this paper. The \planck\/ upper limit at 70 GHz already implies some steepening. 

We note that our results also imply that the observed \g-ray spectrum cannot be due to the bremsstrahlung emission from the population of electrons emitting the radio spectrum. In order to increase the normalization of the relativistic electrons to the $E F_E$ level corresponding to the observed \g-ray peak, we have to strongly decrease the magnetic field strength, to $\ll B_{\rm min}$. This then would imply $m_{\rm e} c^2\gamma_0\gg 1$\,GeV, i.e., the bremsstrahlung peak much above that observed. Alternatively, if we use a value of $\gamma_0$ corresponding to the observed \g-ray peak, the synchrotron spectrum would break at an energy much below that of the highest point of the radio power-law.

Substantial work along the above lines has been done before, see, e.g., \citet{abdo09,abdo10a,abdo10c} and \citet{brandt13}. However, those authors have only obtained some specific sets of the parameters resulting in good fits to their data and did not apply the formalism outlined here. Our new method can be applied to other SNRs for which there is a good measurement of the radio spectrum in a wide frequency range, an estimate of the background matter density, and either a detection or upper limits in the \g-ray range. The presented formalism uses the bremsstrahlung emission of the synchrotron-emitting electrons. However, the method can be easily generalized to include inverse-Compton emission of the same electrons, provided the seed photons for scattering are specified.

\subsection{Comparison to other sources}
\label{comparison}

We compare \src\ to other SNRs detected by the \fermi\/ LAT \citep*{thompson12,hewitt13,brandt15}. \citet{hewitt13} find the \g-ray SNRs fall into two categories: the young ones and ones interacting with a local dense medium (with $n\sim 10^2$ cm$^{-3}$). \src\ appears to belong to the latter category. On the radio vs.\ \g-ray flux diagram of \citet{hewitt13}, \src\ falls below all other identified interacting SNRs. However, this may be due to its relatively large distance; the range of the 1--100 GeV luminosities also shown in \citet{hewitt13} is $\simeq (0.7$--$40)\times 10^{33}$ erg s$^{-1}$, and for $D\simeq 4$\,kpc, the $L(1$--100\,GeV) of \src\ is in the middle of their range. Still, most of the sources shown in \citet{hewitt13} have $L$ above the middle value. The relatively low luminosity of \src\ may then be due to the relatively low density of its molecular medium. 

From \citet{kothes06}, we find the SNR radio indices to be between $\alpha \simeq -0.2$ and $-0.6$, with the index of \src\ being on the hard side. On the other hand, the \g-ray index at $E\geq 1$ GeV of $\simeq 3.2\pm 0.2$ is on the soft side of the observed range of $\Gamma\sim 1.5$--3.5, and overlapping with the range of $\Gamma\simeq 2.1$--3.1 found for SNRs interacting with molecular clouds \citep{brandt15}. The projected size of the SNR, with the $\sim 27'$ total diameter \citep{green14}, corresponds to $\simeq 31(D/4\,{\rm kpc})^{-1}$ pc. The LAT-detected interacting SNRs have diameters of $\simeq$10--50 pc \citep{thompson12,hewitt13}. This comparison also favours the distance of $D\sim 4$ kpc for \src. 

Finally, we note that a number of pulsars (some of them millisecond ones) have spectra similar to that of \src, see \citet{abdo13} and the associated online spectra\footnote{\url{fermi.gsfc.nasa.gov/ssc/data/access/lat/2nd_PSR_catalog/combined_2PC_SED.pdf}}. As some examples, we list PSR J0023+0923, PSR J0101--6422, PSR J0357+3205, PSR J0622+3749, PSR J0908--4913, PSR J1057--5226, PSR J1744--1134, PSR J1833--1034, PSR J2030+3641. However, there is no evidence for the presence of a pulsar in \src, and the similarity may be incidental. 

\section{Conclusions}
\label{conclusions}

We have analysed almost seven years of \fermi\/ LAT data on \src. We have confirmed the detection of high-energy \g-rays from this source by \citet{Malyshev13}, but at a much higher statistical significance of $\simeq 12\sigma$. The detected spectrum covers the 0.1--5 GeV energy range and shows a significant spectral curvature, peaking in $EF_E$ at $\sim$1 GeV. 

We have found that \src\ is most likely located within a molecular cloud, which distance we have estimated as $\sim$4 kpc, confirming \citet{pavlovic13}. We have estimated its density as a function of the source distance. The \g-ray spectrum can be fitted by emission from decay of pions produced in collisions of ions accelerated to relativistic energies with a power-law distribution interacting with the matter of the molecular cloud. 

While the leptonic origin of the \g-ray emission can be ruled out, some bremsstrahlung \g-rays have to be emitted by the relativistic electrons emitting the observed radio flux (via the synchrotron process). Using this, we have developed a method of estimating the magnetic field in the synchrotron-emitting region. It implies a lower limit of $B_{\rm min}\sim 80\,\mu$G if the molecular cloud electron density is $\simeq$30\,cm$^{-3}$ (estimated by us assuming the molecular cloud is spherical with the radius corresponding to $0.15\degr$), or less for a lower density. The magnetic-field estimate is robust for a given background density, almost independent of the shape of the radio spectrum above the measured range.

We have compared \src\/ with other SNRs interacting with surrounding media detected by the LAT, and found the source to be relative similar though with weaker radio and \g-ray emission. 

\section*{Note added in proof}

After this paper was completed and accepted, O. Kargaltsev pointed out to us the existence of a radio galaxy within the field of the SNR, namely WSRT 2011+3600 \citep{taylor96}, with the coordinates of RA 303.40, Dec.\ 36.16, also denoted as 15P21 in \citet{pc90}. Its total radio flux density is $882\pm 20$ mJy at 327 MHz \citep{taylor96} and $238\pm 9$ mJy at 1420 MHz \citep{pc90}. Thus, the source contribution to the integrated SNR radio emission is small, although it is clearly seen as a point source on the radio maps of \citet{kothes06}. \citet{taylor96} gives its size at 327 MHz as $49''\times 23''$. There are also VLA observations at 1438, 1652 and 4886 MHz \citep*{lazio90}. They show a double, FR-II type, source with the two maxima separated by $50''$, and with the southern component about 5 times brighter than the northern one (at all three frequencies), and there is neither an optical nor an X-ray counterpart seen. Thus, the source does not appear to be a blazar. If the source is a typical FR II radio galaxy, its relatively small angular size would suggest it is much more distant than most of radio galaxies detected by \fermi\/ \citep{abdo10b}. We conclude that it is unlikely (though we cannot rule it out) that the source contributes noticeably to the observed GeV emission.

\section*{Acknowledgments}

We thank T. A. Lozinskaya and M. Renaud for valuable discussions. The authors thanks to SFI/HEA Irish Centre for High-End Computing (ICHEC) for the provision of computational facilities and for support. We also thank the referee for valuable comments. This research has been supported in part by the grants AYA2012-39303, SGR2009-811 and iLINK2011-0303. AAZ, JM and RB have been supported in part by the Polish NCN grants 2012/04/M/ST9/00780 and 2013/10/M/ST9/00729.

\appendix
\section{The log-normal/log-parabolic spectrum}
\label{log}

The \fermi\/ LAT analysis software\footnote{\url{fermi.gsfc.nasa.gov/ssc/data/analysis/scitools/source_models.html}} provides a log-parabolic photon-spectrum model for spectral fitting following the form given by \citet{massaro04},
\begin{equation}
{{\rm d}N\over {\rm d}E}=N_0' \left(E\over E_{\rm b}'\right)^{-\alpha-\beta \log_{10}(E/E_{\rm b}')},
\label{logparabola}
\end{equation}
with the four above parameters. However, a parabola has only three independent parameters. In our case, an observed spectrum in the logarithmic space, i.e., $\log_{10}{\rm d}N/ {\rm d}E$, can determine the three coefficients of $\log_{10}^i E$, $i=0,\,1\,2$. The square term has the coefficient of $\beta$, and thus it can be uniquely fitted given an observed spectrum. However, the constant and linear terms are given by combinations of all four variables in equation (\ref{logparabola}). Thus, $N_0'$, $\alpha$, and $E_{\rm b}'$ cannot be separately determined, due to their internal correlation. Indeed, \citet{massaro04} set $E_{\rm b}'$ fixed at 1 keV. 

However, if we keep $E_{\rm b}'$ fixed to some value (as recommended now on the \fermi\/ web page), then it will not have the meaning of the peak of the $E F_E$ ($\propto E^2{\rm d}N/ {\rm d}E$) spectrum, which is a parameter of the log-normal distribution, 
\begin{equation}
{{\rm d}N\over {\rm d}E}=N_0 \left(E\over E_{\rm b}\right)^{-2} 10^{-\beta \log^2_{10}(E/E_{\rm b})}. 
\label{lognormal}
\end{equation}
A similar form of the log-normal distribution was used to fit blazar spectra by, e.g., \citet{landau86}. The values of $N_0$ and $E_{\rm b}$ are given by \citep{massaro04}, 
\begin{equation}
N_0=N_0' 10^{(\alpha^2-4)/(4\beta)},\quad E_{\rm b}=E_{\rm b}' 10^{(2-\alpha)/(2\beta)}.
\label{conversion}
\end{equation}
Their uncertainties then need to be calculated with the propagation of the errors on $\alpha$ and $\beta$. On the other hand, setting $\alpha=2$ and leaving the other three parameters free converts the spectrum of equation (\ref{logparabola}) to the log-normal spectrum in $E F_E$, equation (\ref{lognormal}). In this case, $E_{\rm b}$ does give the peak of $E F_E$. We use this form in our spectral fitting.

\label{lastpage}

\end{document}